\documentclass[twocolumn,floatfix]{revtex4}
\usepackage{graphicx}
\usepackage{amsmath}
\usepackage{amssymb}
\usepackage{bm}

\begin{document}   

\title{Andreev-Bragg reflection from an Amperian superconductor}
\author{P. Baireuther}
\affiliation{Instituut-Lorentz, Universiteit Leiden, P.O. Box 9506, 2300 RA Leiden, The Netherlands}
\author{T. Hyart}
\affiliation{Instituut-Lorentz, Universiteit Leiden, P.O. Box 9506, 2300 RA Leiden, The Netherlands}
\author{B. Tarasinski}
\affiliation{Instituut-Lorentz, Universiteit Leiden, P.O. Box 9506, 2300 RA Leiden, The Netherlands}
\author{C. W. J. Beenakker}
\affiliation{Instituut-Lorentz, Universiteit Leiden, P.O. Box 9506, 2300 RA Leiden, The Netherlands}
\date{March 2015}

\begin{abstract}
We show how an electrical measurement can detect the pairing of electrons on the same side of the Fermi surface (Amperian pairing), recently proposed by Patrick Lee for the pseudogap phase of high-$T_c$ cuprate superconductors. Bragg scattering from the pair-density wave introduces odd multiples of $2k_{\rm F}$ momentum shifts when an electron incident from a normal metal is Andreev-reflected as a hole. These Andreev-Bragg reflections can be detected in a three-terminal device, containing a ballistic Y-junction between normal leads $(1,2)$ and the superconductor. The cross-conductance $dI_1/dV_2$ has the opposite sign for Amperian pairing than it has either in the normal state or for the usual BCS pairing.
\end{abstract} 
 
\maketitle

\textit{Introduction.---} Doped Mott insulators exhibit new symmetry-broken states of matter with coexisting magnetic, charge, and superconducting order \cite{Lee-review,Corboz14, Fradkin-review}. Notable examples of such ``intertwined order'' are superconductors with a pair-density wave (PDW), such that the Cooper pairs acquire a nonzero center-of-mass momentum \cite{Fradkin-review, Himeda02, Han04, Berg07, Raczkowski07, Berg09}. In the first proposals by Fulde, Ferrell, Larkin, and Ovchinnikov (FFLO) the PDW order was induced by an external magnetic field \cite{Fulde60, LO}, but it can appear with preserved time-reversal symmetry in doped Mott insulators (and possibly also in a broader context \cite{Casalbuoni04}). 

In a remarkable recent paper \cite{Lee14}, Patrick Lee has carried this development to its logical endpoint, by proposing PDW order with the maximal $2k_{\rm F}$ Cooper pair momentum. The pairing mechanism comes from the gauge field formulation of the resonating valence bond theory of high-$T_c$ superconductivity \cite{Anderson87, Kotliar88a, Kotliar88b, Lee-review, Senthil09}, where electrons moving in the same direction feel an attractive force analogous to Amp\`{e}re's force between current-carrying wires \cite{Lee07}. Lee has proposed this \textit{Amperian pairing} to explain the diversity of anomalous properties that characterize the pseudogap phase in underdoped cuprate superconductors, including the appearance of Fermi arcs in the quasiparticle spectrum \cite{He11}, charge order with a doping-dependent wave vector \cite{Howald03, Vershinin04, Hanaguri04, Wise08, Ghir12, Chang12, LeTacon14, Comin14, Neto14, Comin14b}, and indications of short-range superconducting order \cite{Wang05, Li10, Dubroka11, Yu14}.

Phase-sensitive experimental tests for Amperian pairing are hindered by phase fluctuations and the nucleation of vortex anti-vortex pairs that are believed to suppress long-range phase coherence \cite{Lee14}. Here we propose to use Andreev reflection as a phase-insensitive probe, which being a local process would not require long-range superconducting order. Earlier studies of the FFLO state have indicated that conductance spectroscopy shows signatures of the nonzero momentum of Cooper pairs \cite{Cui06,Tanaka07,Kac11,Che14}, but these are typically small effects. We find that the extreme $2k_{\rm F}$ momentum transfer upon Andreev reflection from an Amperian superconductor changes \textit{the sign} of the current in a three-terminal configuration, allowing for an well-defined experimental test.

\begin{figure}[!tb] 
\centerline{\includegraphics[width=1.\columnwidth]{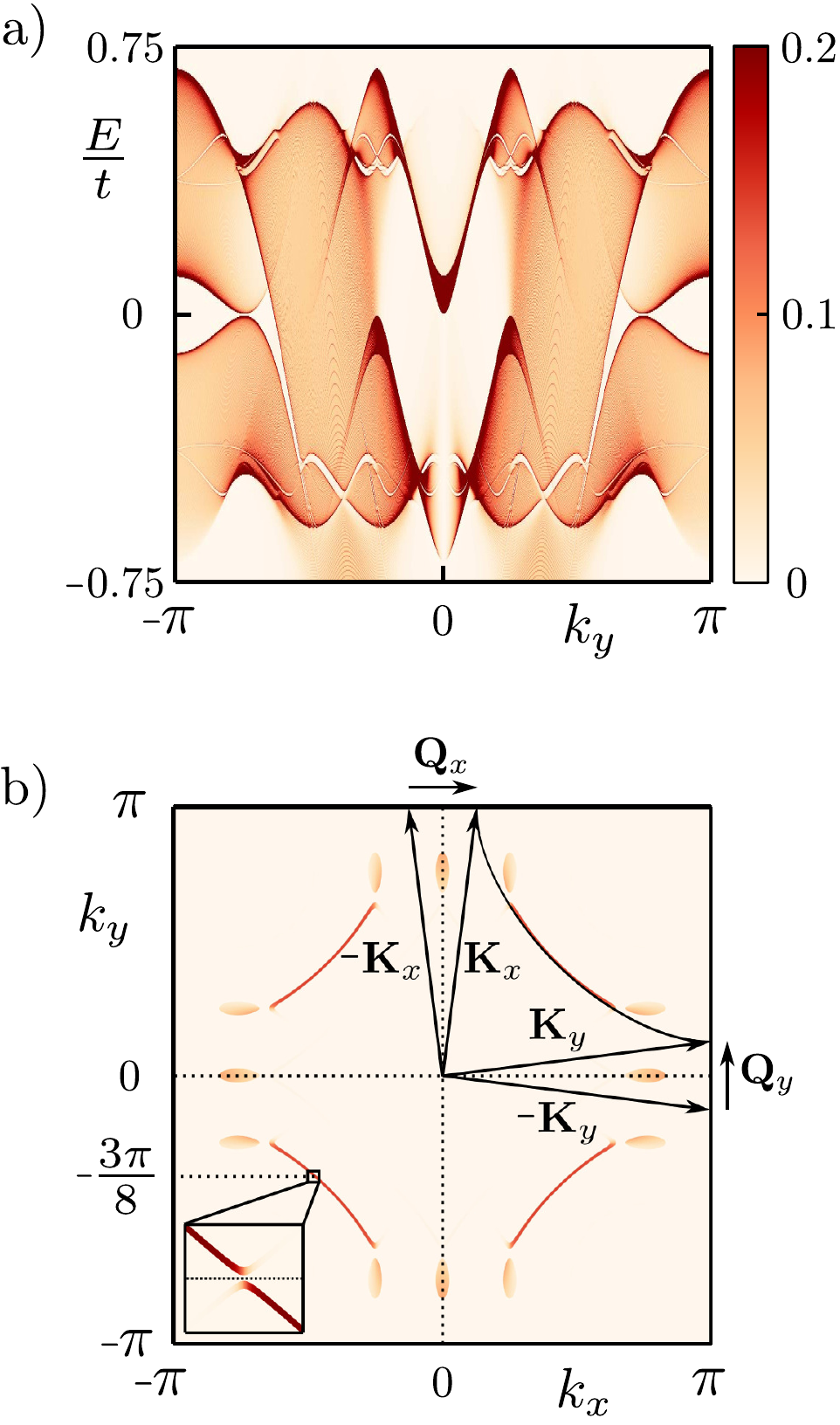}}
    \caption{\textit{a)} Density of states $\rho(E,\bm{k})$ as a function of energy $E$ and momentum $k_y$ integrated over $k_x\in(-\pi,\pi)$. Only the electron contribution is shown (in units of $1/t$), the full density of states also includes the hole contribution $\rho(-E,-\bm{k})$ to ensure particle-hole symmetry. \textit{b)} Density of states as a function of $k_x$ and $k_y$, integrated over a narrow energy interval around the Fermi level $E=0$. The Amperian pairing takes place near the momenta $\pm \bm{K}_x$ and $\pm \bm{K}_y$ where the free-electron Fermi surface crosses the boundary of the first Brillouin zone (black arc and arrows). The PDW with wave vector $\pm \bm{Q}_x$ or $\pm \bm{Q}_y$ has a periodicity of 8 unit cells. This folds the Brillouin zone, but for clarity the figure shows the bands unfolded (extended zone scheme). Inset: Magnified region of the Brillouin zone near $k_y=-3\pi/8$, showing a minigap. 
 }
    \label{fig:dispersion}
\end{figure}
 
 \textit{Model.---} We study the mean-field Hamiltonian
\begin{equation}
  \label{eq:hamiltonian}
  H = \sum_{\bm{k},\sigma} \xi(\bm{k})
  c_{\bm{k}\sigma}^\dagger c_{\bm{k}\sigma} +
  \sum_{i, \bm{k}} \big[ \Delta_{\bm{Q}_i}(\bm{k}) c_{\bm{k}\downarrow}^\dagger c_{\bm{Q}_i-\bm{k},\uparrow}^\dagger +
  {\rm H.c.} \big]
\end{equation}
with square-lattice dispersion
\begin{eqnarray}
\xi(\bm{k})&=&-2t (\cos k_x+\cos k_y)-4t' \cos k_x \cos k_y \nonumber\\ &&-2t'' (\cos 2k_x +\cos 2k_y)-\mu \label{xidef}
\end{eqnarray}
(nearest neighbor hopping energy $t$, chemical potential $\mu$, lattice constant $a\equiv 1$). To make contact with the cuprate superconductor ${\rm Bi}_{2+x}{\rm Sr}_{2-y}{\rm CuO}_{6+\delta}$ (Bi2201), we have also included further-neighbor hopping energies $t'= -0.2\, t$ and $t''= 0.05\, t$ \cite{Comin14}.  

The PDW order parameter $\Delta_{\bm{Q}_i}(\bm{k})$ describes pairing with total momentum $\bm{Q}_i=2\bm{K}_i$ (up to a reciprocal lattice vector) near the points $\bm{K}_i\in\{ \pm \bm{K}_x$, $\pm \bm{K}_y\}$ where the free-fermion Fermi surface crosses the boundary of the first Brillouin zone (see Fig.~\ref{fig:dispersion}\textit{b}). This pairing of electrons on the same side of the Fermi surface defines the Amperian superconductor \cite{Lee14}.
 
Following Lee \cite{Lee14}, we take a phenomenological Gaussian profile (width $k_0$) for the $\bm{k}$-dependence of the order parameter near the momenta $\bm{K}_i$ and their images upon translation by a reciprocal lattice vector $2\pi\bm{j}=2\pi(n,m)$, $n,m\in\mathbb{Z}$:
\begin{equation}
  \label{eq:order_param}
  \Delta_{\bm{Q}_i}(\bm{k}) =
  \frac{\Delta_0}{C} \sum_{\bm{j}} \exp\left( -\frac{|\bm{k}-\bm{K}_{i}- 2\pi \bm{j}|^2}{2k_0^2} \right).
\end{equation}
By choosing the coefficient $C$ such that $\Delta_{\bm{Q}_i}(\bm{K}_i) =  \Delta_0$, the usual BCS order parameter follows in the limit $k_0 \to \infty$ at $\bm{K}_i = 0$. In what follows we set $\Delta_0=0.4\, t$ and $k_0=1.2$~.

We take chemical potential $\mu=-0.75\, t$  corresponding to hole doping fraction $p \approx 0.14$ deep inside the pseudogap phase \cite{Lee-review}. The wave vectors $\bm{Q}_i$ for this doping are $\pm Q_0 \hat{\bm{e}}_x$ and $\pm Q_0 \hat{\bm{e}}_y$ with $Q_0=\pi/4$, corresponding to a PDW periodicity of $8$ square-lattice unit cells.

 \textit{Density of states.---} To prepare for the calculation of the Andreev reflection probability at a normal-superconductor interface, we have first computed the electron density of states in the unbounded superconductor. We use the {\sc kwant} toolbox for all our tight-binding calculations \cite{KWANT}. The result in Fig.~\ref{fig:dispersion} shows the characteristic features of an Amperian superconductor identified by Lee \cite{Lee14}: Fermi arcs and gaps both above and below the Fermi level, in good agreement with experimental data from angle-resolved photoemission spectroscopy (ARPES) \cite{He11, Comin14} and scanning tunneling microscopy \cite{Wise08}. Close inspection reveals that the Fermi arcs are interrupted by a multitude of minigaps (cf. inset of Fig.~\ref{fig:dispersion}\textit{b}), originating from higher order Bragg reflection processes with a momentum shift $\sum_{i} n_i \mathbf{Q}_i$ ($n_i\in\mathbb{Z}$). Lifetime broadening would presumably hide these minigaps from ARPES measurements.

\begin{figure*}[tb]  
 \centerline{\includegraphics[width=1.4 \columnwidth]{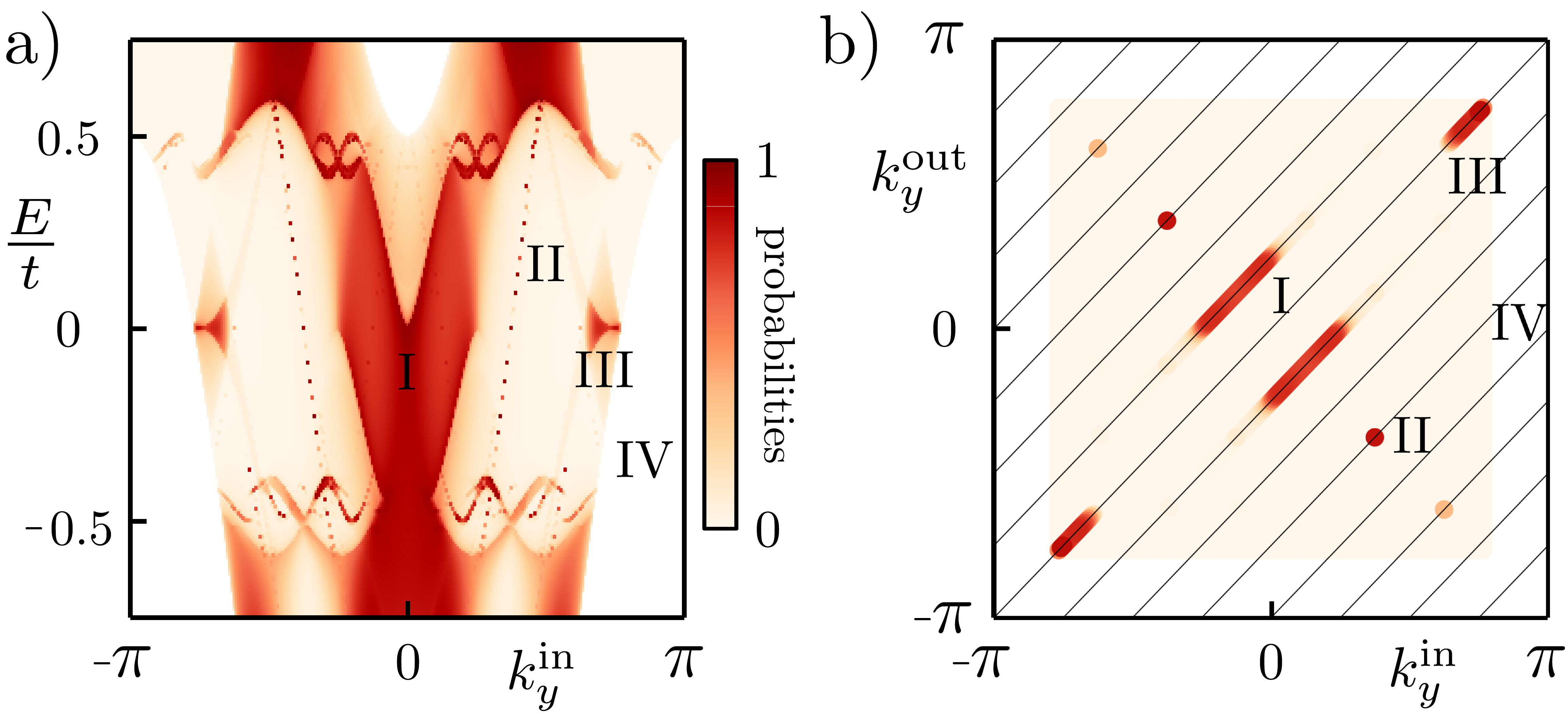}}
    \caption{\textit{a)} Total Andreev reflection probability \eqref{Rtotdef} as a function of energy $E$ and incoming transverse momentum $k_y^{\rm in}$, integrated over the outgoing transverse momenta $k_{y}^{\rm out}$. The regions of non-zero $R_{\rm tot}$ correspond to the gapped regions in Fig.~\ref{fig:dispersion}. \textit{b)} Andreev reflection probability at $E=0$ as a function of incoming and outgoing transverse momenta. The diagonal black lines correspond to momentum shifts $k_y^{\rm out}=k_y^{\rm in}+n Q_0$, $n \in \mathbb{Z}$. Region I around $E=0$ and $k_y^{\rm in}=0$ exhibits Andreev-Bragg reflection with transverse momentum shift $Q_0$ between the incoming electron and the outgoing hole. Region II shows transverse momentum shifts of $3Q_0$. (The resonance lines labeled II in panel \textit{a} are interrupted by the finite $k_y$ resolution.) Region III supports Andreev retroreflection (without transverse momentum shift), because the  periodic modulation of the order parameter is perpendicular to the normal-superconductor interface. Region IV (white) has no incoming electron modes from the normal metal.}
    \label{fig:angle_resolved_AR}
\end{figure*}
 
\textit{Andreev-Bragg reflection.---} We now introduce an interface with a normal metal along the line $x=0$, extended in the $y$-direction over $256$ lattice sites with periodic boundary conditions. The Amperian superconductor is at $x>0$, with Hamiltonian \eqref{eq:hamiltonian}, while for the normal metal at $x<0$ we take a nearest-neighbor tight-binding Hamiltonian (same $a$ and $t$, $\mu=-0.5\,t$, $\Delta_0=0$, $t'=t''=0$). We adopt the so-called maximum contact boundary conditions of Ref.\ \onlinecite{Tanaka07}, whereby the periodic modulation of the order parameter in the $x$-direction has a maximum at the $x=0$ interface.

We inject an electron with energy $E$ and transverse momentum $k_y^{\rm in}$ from the normal metal towards the superconductor and calculate the probability $R(E,k_y^{\rm in},k_y^{\rm out})$ for Andreev reflection as a hole with transverse momentum $k_{y}^{\rm out}$. Fig.~\ref{fig:angle_resolved_AR}\textit{a} shows the total Andreev reflection probability
\begin{equation}
R_{\rm tot}(E,k_y^{\rm in})=\int_{-\pi}^{\pi}dk_y^{\rm out}\,R(E,k_y^{\rm in},k_y^{\rm out}),\label{Rtotdef}
\end{equation}
while Fig.~\ref{fig:angle_resolved_AR}\textit{b} shows how the probability at the Fermi-level $R(0,k_y^{\rm in},k_y^{\rm out})$ varies as a function of incoming and outgoing transverse momenta.
  
As can be seen in Fig.~\ref{fig:angle_resolved_AR}\textit{a}, there are distinct regions I, II, III of nonzero $R_{\rm tot}$, each with a gapped density of states (cf.\ Fig.~\ref{fig:dispersion}\textit{a}). The corresponding Andreev reflection processes can be understood by recalling that the Amperian superconductor is described by a bi-directional (checkerboard) modulation of the order parameter with periodicity $2\pi/Q_0$ along both the $x$- and $y$-directions. Since the interface is parallel to the $y$-direction the modulation along $x$ gives rise to usual Andreev retroreflection without a momentum shift (region III). In contrast, the modulation along $y$ produces Andreev-Bragg reflection with transverse momentum shift $n Q_0$ ($n\in 2\mathbb{Z}+1$). The order $n=\pm 1$ and $n=\pm 3$ processes are visible in Fig.~\ref{fig:angle_resolved_AR}\textit{b}, in regions I and II, respectively. Momentum shifts at even multiples of $Q_0$ do not appear, because these produce only normal reflection (without electron-to-hole conversion).

\begin{figure}[tb]
\centerline{\includegraphics[width=1\columnwidth]{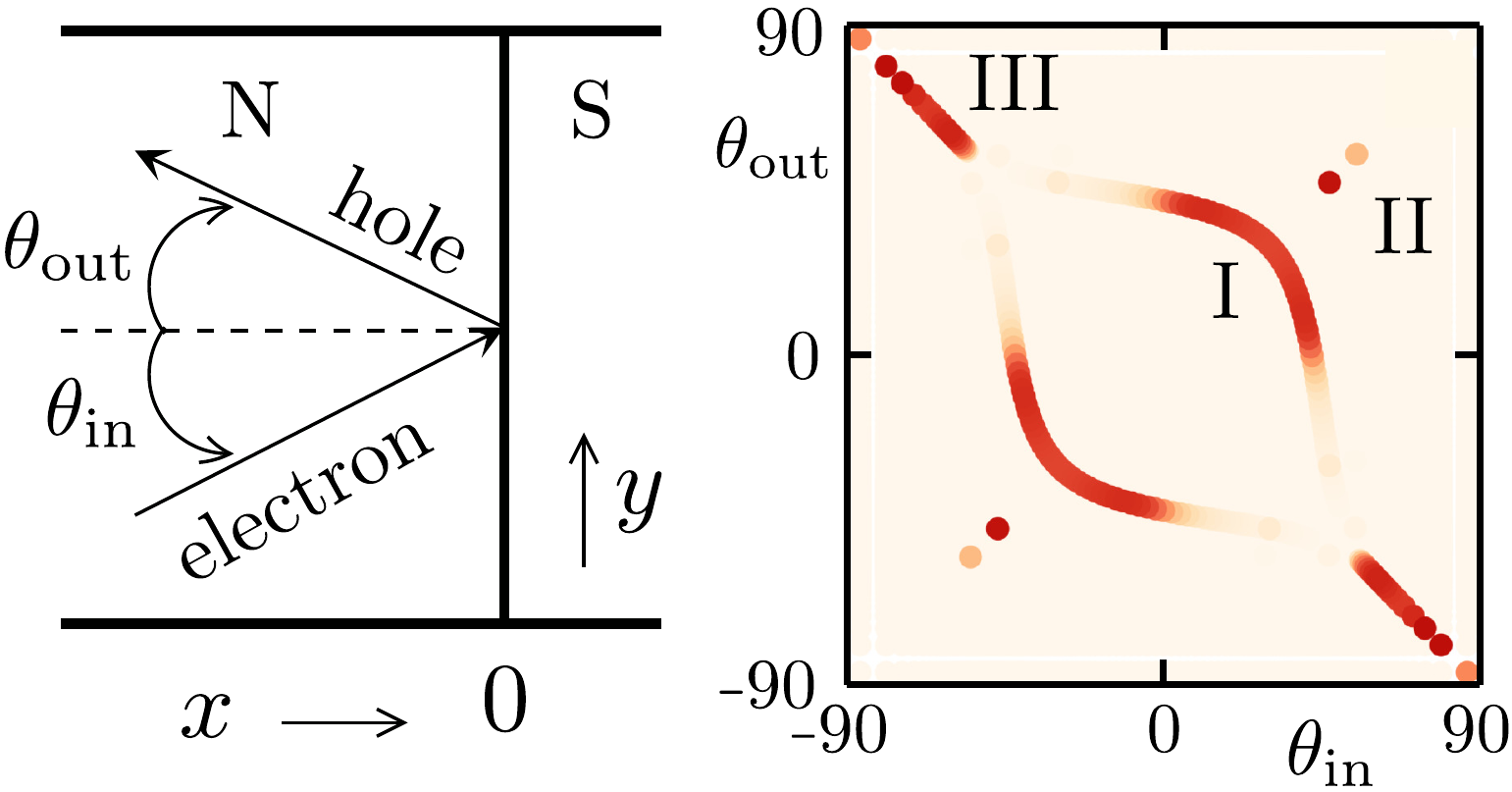}}
    \caption{Right panel: Same as Fig.~\ref{fig:angle_resolved_AR}\textit{b}, but now as a function of angles $\theta$ of velocities (measured in degrees) rather than transverse momentum $k_y$. The left panel indicates the geometry of the normal-superconductor (NS) interface, with retroreflection corresponding to $\theta_{\rm out}=-\theta_{\rm in}$ and specular reflection to $\theta_{\rm out}=\theta_{\rm in}$.}
    \label{fig:double_angle_resolved_AR}
\end{figure}

The angular dependence in real space of the Andreev reflection processes of type I, II, and III is shown in Fig.~\ref{fig:double_angle_resolved_AR}.  The directionality of Andreev-Bragg reflection is centered around specular reflection ($\theta_{\rm in}=\theta_{\rm out}$), with a broad spread of angles for the first-order Bragg shift (type I) and a narrow collimation for higher orders (type II). The conventional Andreev retroreflection (type III, $\theta_{\rm in}=-\theta_{\rm out}$) appears only near grazing incidence.

\begin{figure}[tb]  
\centerline{\includegraphics[width=0.9\columnwidth]{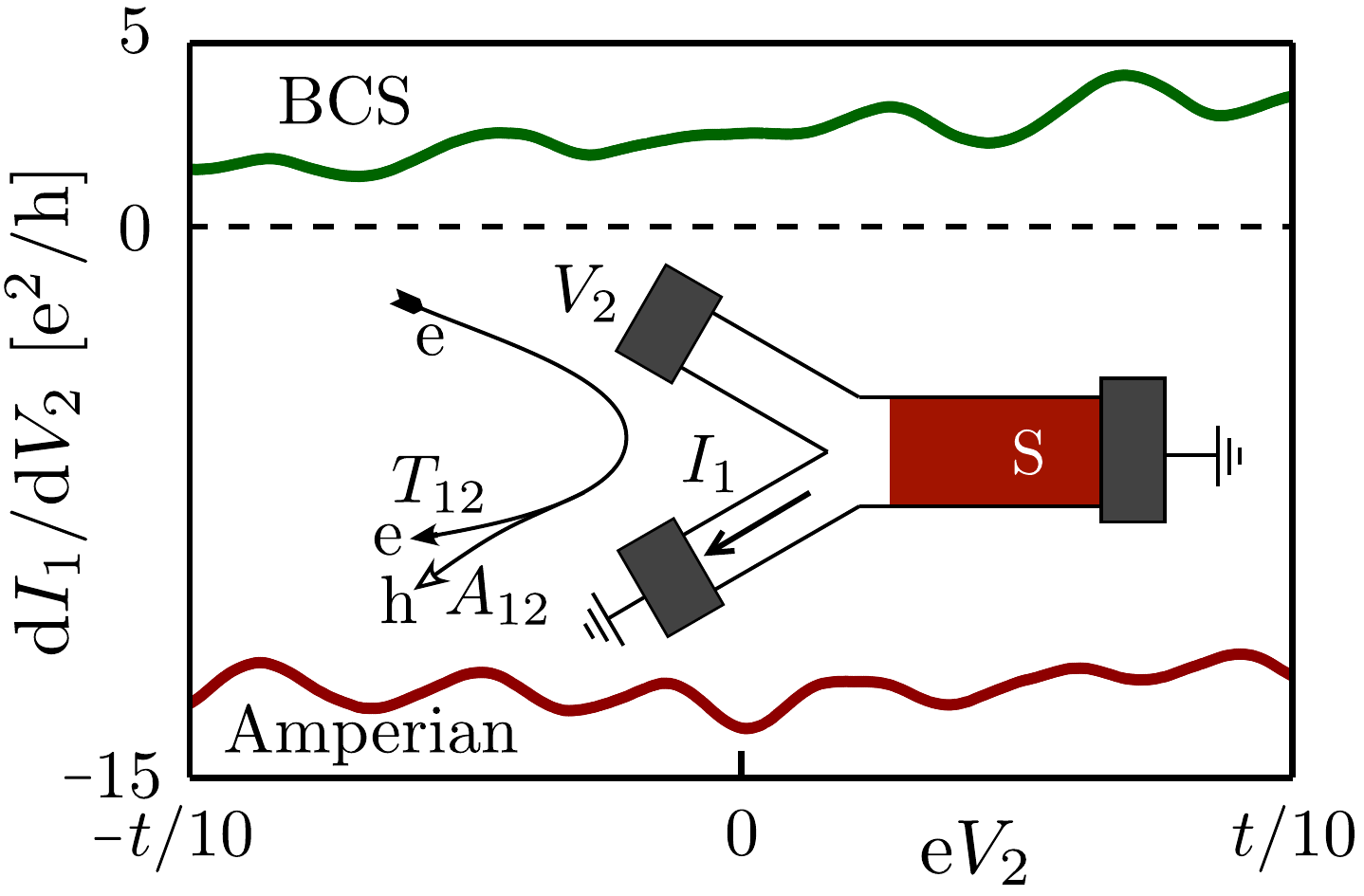}}
    \caption{Computer simulation of the differential cross-conductance for the Y-junction geometry shown in the inset. The sign of the cross-conductance $dI_1/dV_2\propto T_{12}-A_{12}$ distinguishes Amperian pairing ($A_{12}>T_{12}$) from BCS pairing ($T_{12}>A_{12}$).}
    \label{fig:yjunction}
\end{figure}

{\it Method of detection---} Electrical detection of momentum transfer upon Andreev reflection has been proposed in the context of FFLO superconductors, notably using a magnetic-flux controlled interferometer \cite{Che14}. (A similar Aharonov-Bohm interferometer has been proposed \cite{Sch12} to detect specular Andreev reflection in graphene \cite{Bee06}.) Here we investigate an alternative electrical method of detection of Andreev-Bragg reflection that relies on ballistic transport, but does not require long-range phase coherence and might therefore be more easily realized. 

We consider the three-terminal Y-junction of Fig.~\ref{fig:yjunction} (inset), similar to geometries considered for the detection of ``Cooper pair splitting'' \cite{Hof09}. The current $I_1$ flowing into the grounded normal-metal contact 1 is measured while the other normal-metal contact 2 is biased at voltage $V_2$. (The superconductor is also grounded \cite{note}.) The differential cross-conductance $dI_1/dV_2$ is expressed by a three-terminal variation of the Blonder-Tinkham-Klapwijk formula \cite{BTK82},
\begin{equation}
  \frac{dI_1}{dV_2} = \frac{2 e^2}{h} \int_{-\infty}^\infty dE\, \big[T_{12}(E)-A_{12}(E)\big]\frac{df(E-eV_2)}{-dE},\label{BTK}
\end{equation}
in terms of the probabilities (summed over all transverse modes) for an electron to be transmitted from contact 2 into contact 1, either as an electron (normal transmission probability $T_{12}$) or as a hole (crossed Andreev reflection probability $A_{12}$). The probabilities are integrated over energy $E$, weighted by the derivative of the Fermi distribution $f(E)=(e^{\beta E}+1)^{-1}$. 

We have performed computer simulations to determine whether such a device has sufficient angular resolution to distinguish Andreev-Bragg reflection from the usual retroreflection. We took a $60^\circ$ angle between the two normal-metal leads, each 146 lattice constants wide, with open boundary conditions. The conductance was calculated from Eq.\ \eqref{BTK} at a temperature of $0.01\,\Delta_0$. We compared Amperian pairing with BCS pairing, keeping all other parameters of the tight-binding Hamiltonian the same.
 
The results plotted in Fig.~\ref{fig:yjunction} demonstrate that for a large range of voltages the differential cross-conductance is negative in the Amperian case ($A_{12}>T_{12}$, because Andreev-Bragg reflection dominates) and positive in the BCS case ($T_{12}>A_{12}$, because retroreflection dominates). Notice that an entirely normal system would have $A_{12}\equiv 0$, hence $dI_1/dV_2>0$ --- so the negative cross-conductance can only originate from Andreev reflection.

\begin{figure}[tb]  
\centerline{\includegraphics[width=0.8\columnwidth]{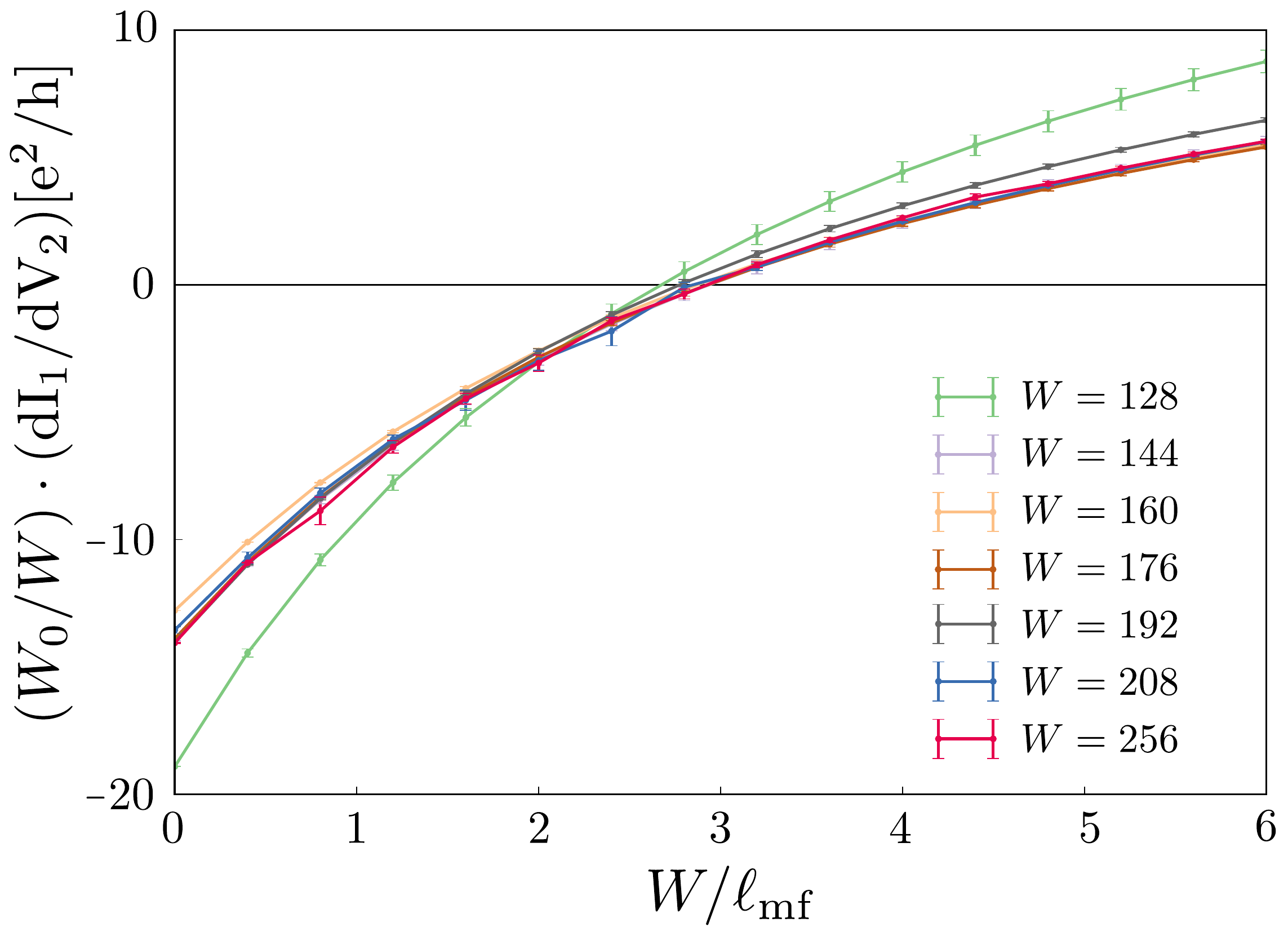}}
    \caption{Effect of disorder on the cross-conductance of Fig.\ \ref{fig:yjunction}, for Amperian pairing at $V_2=0$. The data is averaged over four disorder realizations (error bars show standard error of the mean). Data collapse for different system widths $W$ is achieved by rescaling the conductance by a factor $W_0/W$, with $W_0=256$. The negative cross-conductance persists for $W/\ell_{\rm mf}\lesssim 2$.}
    \label{fig:disorder}  
\end{figure}

{\it Effects of disorder and interface barrier.---}
Because the negative cross-conductance is a ballistic effect, strong impurity scattering will obscure it, but the sign change should persist if the mean free path $\ell_{\rm mf}$ is not much smaller than the width $W$ of the junction. To confirm this, we model electrostatic disorder by a random on-site energy with a Gaussian distribution with variance $\sigma^2$, resulting in $\ell_{\rm mf}= \hbar v_{\rm F}(2\pi N_0\sigma^2)^{-1}\approx 0.9(t/\sigma)^2$. (We have used that the metallic part of the Y-junction has a nearly circular Fermi surface, with density of states per spin $N_0$.) Results shown in Fig.\ \ref{fig:disorder} confirm our expectation.
%assuming also that disorder does not prevent the formation of the pair-density wave \cite{Berg09}.

\begin{figure}[tb]   
\centerline{\includegraphics[width=0.8\columnwidth]{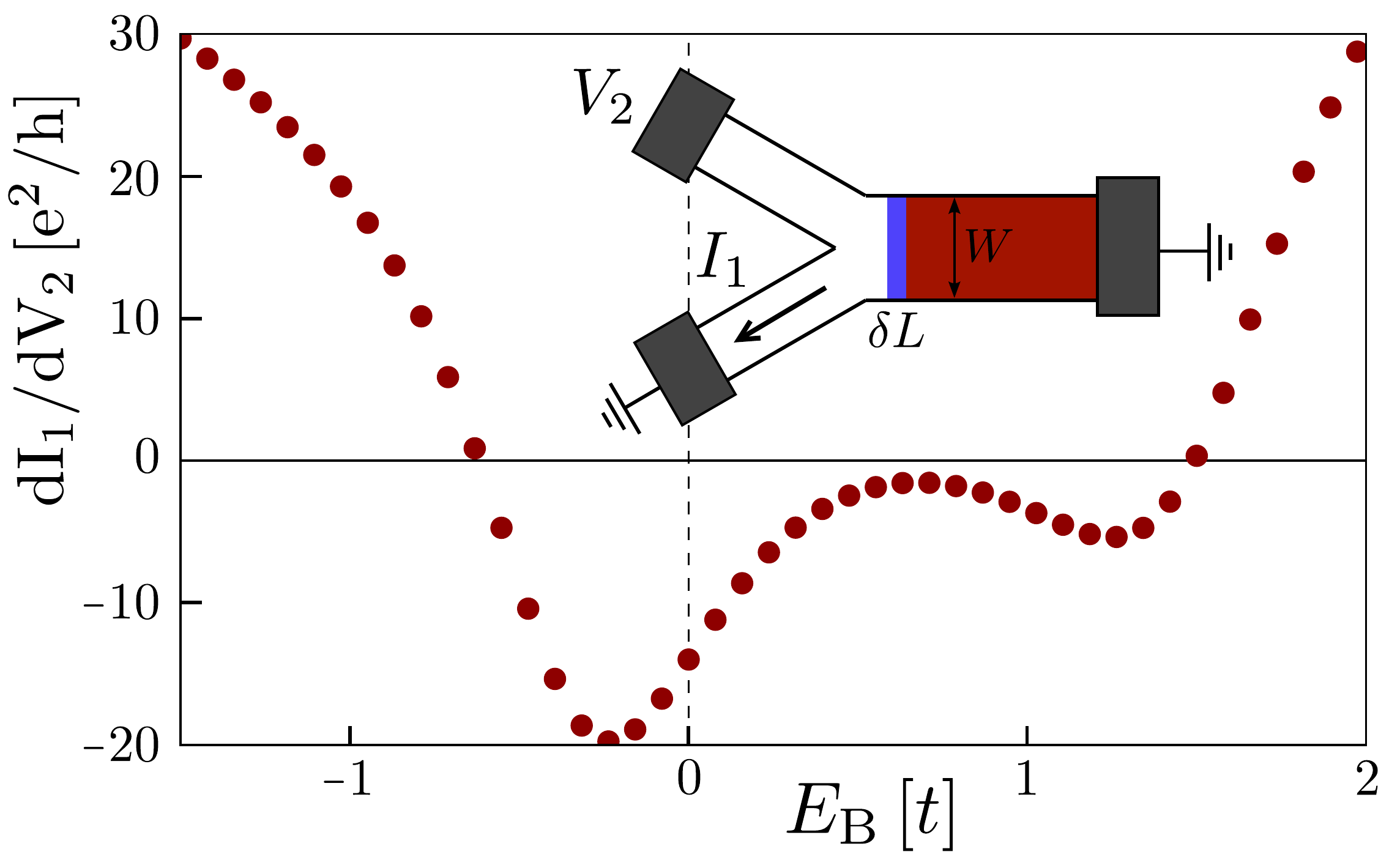}}
    \caption{Effect of an interface barrier (height $E_{\rm B}$, length $\delta L=3$, width $W=256$) on the cross-conductance of Fig.\ \ref{fig:yjunction}, for Amperian pairing at $V_2=0$.}
    \label{fig:barrier}
\end{figure}

Another detrimental effect is the presence of a barrier at the interface with the superconductor, since this would suppress Andreev reflection in favor of normal reflection. The computer simulation of Fig.\ \ref{fig:barrier} shows that the effect of a tunnel barrier has a significant electron-hole asymmetry, but eventually for barrier heights $|E_{\rm B}|\gtrsim t$ the negative cross-conductance disappears. A high quality interface is therefore needed.

{\it Conclusion.---} We have shown that Andreev reflection from an Amperian superconductor involves a transverse momentum transfer of odd multiples of $Q_0=2k_F$, because of Bragg scattering from the pair-density wave. Computer simulations show that this Andreev-Bragg reflection can be detected in a Y-junction, through a sign change of the differential cross-conductance. Long-range phase coherence is likely to be absent in the Amperian superconductor \cite{Lee14}, but since Andreev reflection is a local process we expect the predicted experimental signature of the $2k_{\rm F}$ pairing to be robust and accessible.

The experimental signature of Andreev-Bragg reflection proposed here in the context of cuprate superconductors may be of use in other contexts as well. The pair-density waves predicted in Mott insulators \cite{Fradkin-review} have crystal momentum on the order of the inverse lattice constant, and might therefore be detected via a negative cross-conductance in the geometry of  Fig.\ \ref{fig:yjunction}. Ultracold fermionic atoms in a two-dimensional optical lattice provide an altogether different realization of the Hubbard model \cite{Schneider08}. Thus, these systems may also support PDW states, which can potentially be detected via Andreev-Bragg reflections. An energy-resolved scattering experiment is challenging in that context, but in the light of the recent observation of conductance quantization in a cold atom setup \cite{Krinner14}, the analogue of negative cross-conductance may become observable as well.

{\it Acknowledgments.---} This research was supported by the Foundation for Fundamental Research on Matter (FOM), the Netherlands Organization for Scientific Research (NWO/OCW), and an ERC Synergy Grant.

\end{document}